\documentclass[runningheads]{llncs}
\usepackage{graphicx}
\begin{document}
%
\title{Domain Specific Transporter Framework to Detect Fractures in Ultrasound}
\titlerunning{Domain Specific Transporter Framework to Detect Fractures in Ultrasound}
%
\author{Arpan Tripathi\inst{1} \and
Abhilash Rakkunedeth\inst{2} \and
Mahesh Raveendranatha Panicker\inst{1} \and
Jack Zhang\inst{2} \and
Naveenjyote Boora\inst{2} \and
Jacob Jaremko\inst{2}}

%
\authorrunning{Tripathi et al.}
%
\institute{Indian Institute of Technology Palakkad, India \and
University of Alberta,Canada\\}
%
\maketitle              
\begin{abstract}
Ultrasound examination for detecting fractures is ideally suited for Emergency Departments (ED) as it is relatively fast, safe (from ionizing radiation), has dynamic imaging capability and is easily portable. High interobserver variability in manual assessment of ultrasound scans has piqued research interest in automatic assessment techniques using Deep Learning (DL). Most DL techniques are supervised and are trained on large numbers of labeled data which is expensive and requires many hours of careful annotation by experts. In this paper, we propose an unsupervised, domain specific transporter framework to identify relevant keypoints from wrist ultrasound scans. Our framework provides a concise geometric representation highlighting regions with high structural variation in a 3D ultrasound (3DUS) sequence. We also incorporate domain specific information represented by instantaneous local phase (LP) which detects bone features from 3DUS. We validate the technique on 3DUS videos obtained from 30 subjects. Each ultrasound scan was independently  assessed by three readers to identify fractures along with the corresponding x-ray. Saliency of keypoints detected in the image\ are compared against manual assessment based on distance from relevant features.The transporter neural network was able to accurately detect 180 out of 250 bone regions sampled from wrist ultrasound videos. We expect this technique to increase the applicability of ultrasound in fracture detection. 

\end{abstract}
\section{Introduction}

Fractures at the wrist and elbow are common in children and are usually diagnosed in Emergency Departments (ED) using X-ray examination \cite{ref_1}, \cite{ref_2}. This involves radiation exposure which is highly undesirable considering that fractures are not found in many cases. Ultrasound examination is a much safer alternative that is highly sensitive to cortical disruption \cite{ref_3}, \cite{ref_4} which could potentially reveal fractures earlier than x-rays. Various studies have validated the  feasibility of using ultrasound examination for diagnosing pediatric distal radial fractures \cite{ref_3}–\cite{ref_7}. Meta-analysis on 1204 patients has reported 97 percent sensitivity and 95 percent for ultrasound examination \cite{ref_8}.

\paragraph{}
However in clinical practice, acquiring high quality wrist ultrasound scans requires many hours of training which is challenging in ED. Often scanning is performed by non-experts resulting in low quality images which are  hard to interpret. 3D ultrasound (3DUS) addresses this limitation in part by making it easier for novice users to acquire images of adequate quality when compared to 2D ultrasound \cite{ref_9}. It  also depicts the fracture more completely  by including regions on either side of a fracture which could be useful for treatment planning of more complex anatomy. However,manual assessment of both 3DUS and 2DUS are still subjective resulting in high variability between readers. Automated assessment would eliminate such variation and lead to wider use of ultrasound in fracture detection.

\paragraph{}
The presence of noise artefacts and blurred image boundaries make automatic interpretation of ultrasound more difficult when compared to MRI, CT and X-ray. Bony structures like wrist images are even harder due to the effect of beamwidth \cite{ref_10} and shadowing which significantly alters the pixel-intensities around the bone. Compared to intensity based approaches, instantaneous local phase (LP) generally provides more robust information on the underlying structures in noisy ultrasound images \cite{ref_11}. Local phase information alone might not be sufficient to localize the bone as there could be similar echogenic regions in close proximity. LP filtering would also lose some of the useful information in the original b-mode image. Recently, Convolutional neural network(CNN) models that fuse LP filtered images with the original b-mode image have been used for bone segmentation \cite{ref_12}.

\paragraph{}
A major limitation of supervised models like CNNs is the need for a large number of labeled data which is tedious, time consuming and expensive since it requires medical expertise. Even in cases where labeled data is available there could be variability between manual annotations leading to uncertainty in ground truth. Over recent years, semi-supervised and unsupervised models have gained prominence in image analysis as they significantly reduce the dependence on labeled data. These models exploit spatial \cite{ref_13} and temporal \cite{ref_14} dependencies in the data to generate low dimensional representations. In video image analysis, unsupervised learning based on automatically detected keypoints offers an explainable framework to learn concise geometric representations \cite{ref_15,ref_16}. The transporter framework proposed in \cite{ref_15,ref_16} automatically identifies keypoints in a video sequence by transporting features from a target frame. The loss function minimizes the Mean Square Error (MSE) between the reconstructed feature map and target image.

\paragraph{}
In this paper, we propose an unsupervised  transporter neural network to detect keypoints from wrist ultrasound images. A key contribution in our approach is integration of a bone probability map that allows the model to focus on specific regions in the image filtered by local phase at multiple scales. We use a new Acoustic Feature Fusion Convolutional Neural Network (FF-CNN) to generate features relevant to ultrasound based on the bone probability map.This would be first work to use unsupervised keypoint detection in ultrasound sweeps.

\section{Methods}
\begin{figure}
\includegraphics[width=\textwidth]{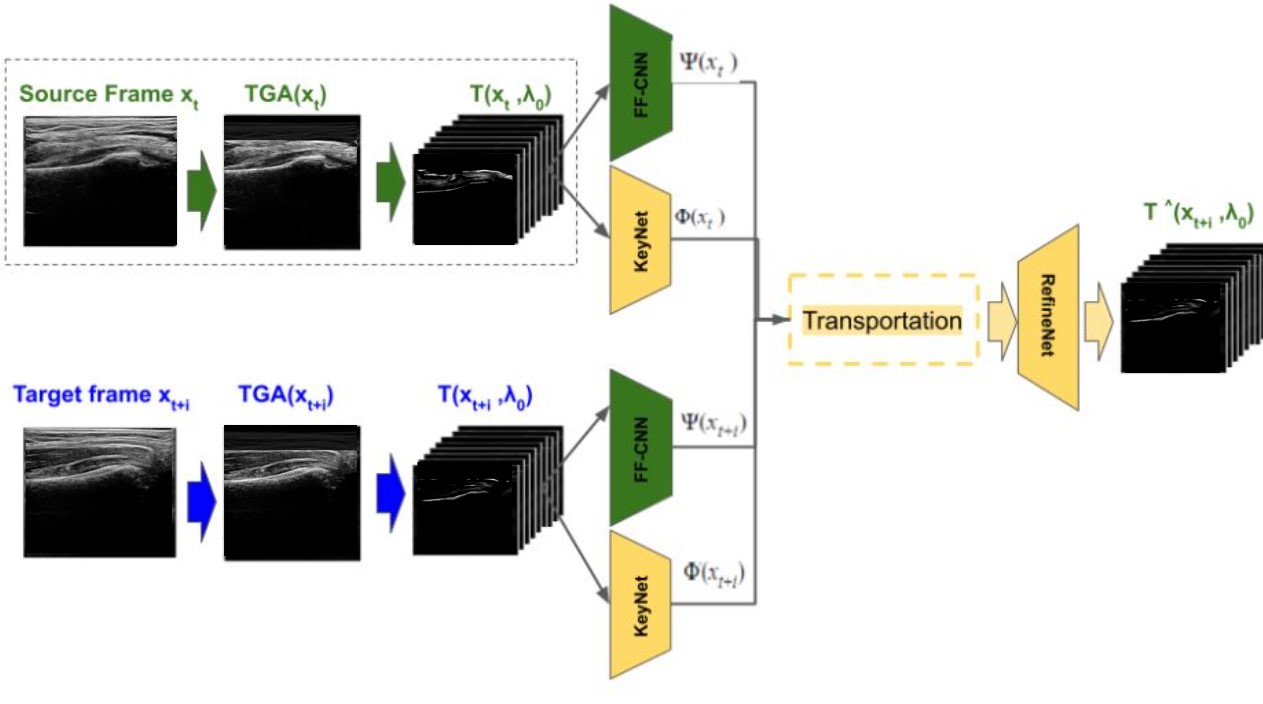}
\setlength{\belowcaptionskip}{-20pt}
\caption{The transporter neural network framework enables unsupervised learning using a pair of frames that are TGA compensated and decomposed into LP filtered channels. The feature map and point map for each frame are generated using FF-CNN and KeyNet. Features from the target feature map are copied at keypoint locations in the source. RefineNet performs inpainting on this feature map to reconstruct the local phase components of the target frame.} \label{fig1}
\end{figure}

An overview of the proposed neural network architecture is shown in Figure \ref{fig1}. For each frame $x_{t}$ in the 3DUS video, the Acoustic FF-CNN and KeyNet generate a feature-map ($\Psi\left(x_{t}\right)$) and point-map ($\Phi\left(x_{t}\right)$) respectively. Using the transportation technique described in \cite{ref_15,ref_16}, features from the feature map of a target frame ($\Psi\left(x_{t+i}\right)$) at their corresponding keypoint locations ($\Phi\left(x_{t+1}\right)$) are embedded into the feature map ($\Psi\left(x_{t}\right)$) at keypoint locations described in ($\Phi\left(x_{t}\right)$). A generative model (RefineNet) uses this augmented feature map to reconstructed the target frame $x_{t+i}$.

\subsection{Time Gain Attenuation (TGA)}
We use TGA to pre-process the image and suppress unwanted bright regions which generally occur due to  acoustic reflection. TGA applied to image $x$ can be represented as $x_{TGA}=\gamma . x$ where $\gamma$ represents a depth dependent decay mask whose value at depth $d$, i.e, $\gamma(d)=1-e^{-a \times d} / \max \left(e^{-a \times d}\right)$ depends on an exponential attenuation factor $a$. 

\subsection{Bone Probability Map}
We extract local phase information from the TGA compensated image using a Gabor filter bank $G\left(x, \lambda_{0}\right)$ where $\lambda_{0}$ is a scaling parameter that can be varied  as to capture image features at different resolutions.

\begin{equation}
G(\omega)=\exp \left(\frac{-\left(\log \left(\frac{\mid\omega \mid}{\omega_{0}}\right)\right)^{2}}{2\left(\log \left(\sigma_{0}\right)\right)^{2}}\right) \textrm{, where } \omega_{0}=\frac{2 \pi}{\lambda_{0}}
\end{equation}

As shown in Figure \ref{fig1} the output of the Gabor filter bank analyzes the image under various frequency bands. We use the monogenic signal analysis described in \cite{ref_11} to generate the bone probability map. In monogenic analysis \cite{ref_11} we model an image $I(x, y): I \in R^{2}$ using a combination of amplitude $A(x,y)$  and phase  as 
\begin{equation}
(x, y)=A(x, y) \times \cos (\theta)
\end{equation}

The corresponding local phase filtered image can be written in tensor representation using symmetric features ($T_{even}$), asymmetric features ($T_{odd}$) and instantaneous phase $\varphi$ as :
\begin{equation}
L P T=\sqrt{T_{even}^{2}+T_{odd}^{2}} \times \cos (\varphi)    
\end{equation}

Tensors of symmetric and asymmetric features ($T_{even}$ and $T_{odd}$)are computed using Hessian $H$, Gradient $\nabla$, and Laplacian $\nabla^{2}$ operations as shown below:

\begin{equation}
\begin{array}{l}
T_{even}=\left[H\left(I_{B P}(x, y)\right)\right]\left[H\left(I_{B P}(x, y)\right)\right]^{T} \\
T_{odd}=-0.5 \times\left[\nabla\left(I_{B P}(x, y)\right)\right]\left[\nabla \nabla^{2}\left(I_{B P}(x, y)\right)\right]^{T}+\\\left[\nabla \nabla^{2}\left(I_{B P}(x, y)\right)\right]\left[\nabla\left(I_{B P}(x, y)\right)\right]^{T}
\end{array}   
\end{equation}

We calculate three monogenic signals $M_{1}$ , $M_{2}$ and $M_{3}$ by applying the Riesz transform on LPT as described in \cite{ref_17} . Using these monogenic signals we define the Local Phase($LP(x)$), feature symmetry ($FS(x)$)  and pixel wise Integrated Backscatter map ($IBS(i,j)$) :
\begin{equation}
\begin{array}{l}
LP(x)=1-\arctan \left(\frac{\sqrt{M_{2}^{2}+M_{3}^{2}}}{M_{1}}\right) \\
FS(x)=\frac{\max \left(T_{even}-T_{odd}-\tau\right)}{M_{1}^{2}+M_{2}^{2}+M_{3}^{2}}
\end{array}    
\end{equation}

$IBS(i, j)=\sum_{k=1}^{i} I^{2}(i,j)$ represents the integrated back scatterer map along the row direction. Finally we combine these to generate a bone probability map $T\left(x, \lambda_{0}\right)$:
\begin{equation}
T\left(x, \lambda_{0}\right)=LP(x) \times FS(x) \times(1-IBS(x))
\end{equation}

\subsection{Acoustic FF-CNN}
The Acoustic FF-CNN combines the multi-channel images generated with the bone probability maps at different scales as in $T\left(x, \lambda_{0}\right)$. Since the input is the multi-channel multi resolution images, the acoustic FF-CNN ($\psi(x_t)$) could be considered as a multi-resolution encoding network.   

\subsection{KeyNet}
The KeyNet also uses the same input as the FF-CNN and  generates a gaussian heat map of keypoints.

\subsection{RefineNet}
The Refine net is a  generative model that reconstructs the target frame based on the transported \cite{ref_15,ref_16} feature map by inpainting regions in the neighbourhood of the keypoints.

\subsection{Ultrasound Scanning}

We prospectively collected ultrasound scans from 30 children aged less than 17 years presenting at an emergency department with suspected unilateral fracture at the wrist. We collected single 2D images, sweeps and 3DUS images from both the affected and unaffected wrist along with conventional x-ray examination which was used to obtain gold standard diagnosis. 

\paragraph{}
Images were acquired on a Philips iU22 machine (using a 13-MHz 13VL5 probe for 3DUS) with the child seated in a neutral position. During examination, the injured wrist  was scanned on the dorsal (DS) and volar (VO) surfaces in both the sagittal and axial orientation resulting in four 3D scans of each wrist (DS sagittal,DS axial, VO sagittal and VO axial). While acquiring the 3DUS image the sonographer centered the view on the distal end of the radius in the different orientations. Each sweep was of 3.2 seconds duration through a range of +/-$15^{o}$ to with  ultrasound slices of 0.2mm. As a baseline, the same scanning protocol was followed for the unaffected wrist as well.  

\subsection{Training Details}
The Transporter architecture was implemented in PyTorch v1.7.1 and trained for 100 epochs with Adam optimizer,learning rate = 0.001 decaying by 0.95/10 epochs, batch size = 16, training set = 1024 pairs of images and validation set = 512 for detecting 10 keypoints. The remaining hyper parameters were adopted from \cite{ref_15}. Samples were drawn from grayscale Wrist Ultrasound videos resized to 256x256 resolution at ~25 FPS as pairs of frames separated by i=4 frames( refer Figure \ref{fig1}).
\paragraph{}
During evaluation, each frame is forward passed through the KeyNet and every $j$th output channel up-sampled from 64x64 resolution to 256x256 in order to plot the $j$th keypoint on the input frame.

\section{Results}
We validated our technique on 56 3DUS videos (containing 256 frames each) of ultrasound wrist. 22 / 30 scans had fractures visible in at least one view. Three human readers with  varying years of expertise  examined  each ultrasound video along with the corresponding x-ray and reported fractures. Ground truth was established based on consensus between the human readers.

\subsection{Bone Probability Map}
\begin{figure}
\includegraphics[width=\textwidth]{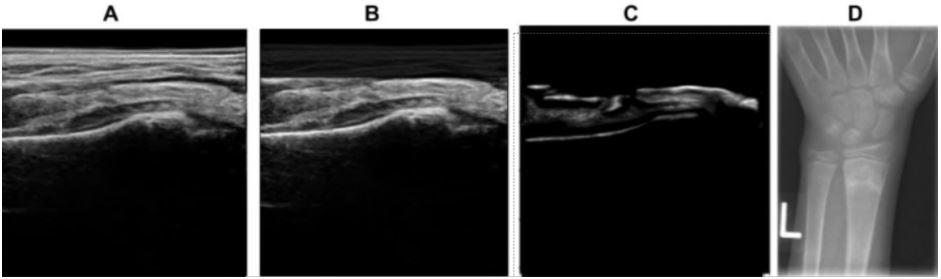}
\setlength{\belowcaptionskip}{-18pt}
\caption{Examples of the original image (A), TGA compensated image (B) and the corresponding bone probability map (C)  and the corresponding x-ray (D). TGA compensation excludes bright patches seen in the top portion of images and correctly identifies the bony structures.} 
\label{fig2}
\end{figure}

TGA compensation was applied to all frames before generating the bone probability maps (refer Figure \ref{fig2}C). It can be seen from the figure that the bone probability map accurately captures features of the bone that are involved in the fracture.Location of the fracture as seen in ultrasound, along with the corresponding x-ray is shown in Figures \ref{fig2}A and \ref{fig2}D.

\subsection{Keypoint Detection}
In affected as well as unaffected wrists the neural networks correctly identified the top portion of the bone with multiple keypoints (refer Figure \ref{fig2}). In affected wrists (Figure \ref{fig2}B) the network was able to track keypoints near the fracture. We manually selected a tight rectangular region of interest around the bone and the neural network correctly identified key points within this region in 180 / 250 ultrasound frames.

\begin{figure}
\includegraphics[width=0.5\textwidth]{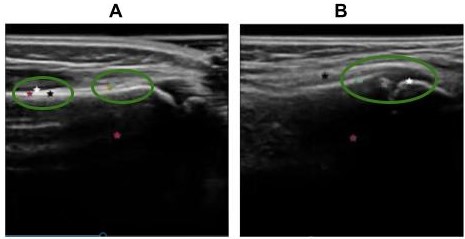}
\setlength{\belowcaptionskip}{-18pt}
\caption{Example of slices in which the neural network correctly identifies keypoints along the boundary of the bone. A: Intact bone imaged in the Volar Sagittal(VS) view B: Ultrasound frame capturing the fracture in the VS view. Note that the network identified points near the bone discontinuity indicating fracture.} \label{fig3}
\end{figure}


\subsection{Ablation Studies}
In order to determine the optimal parameters for various network components, we performed ablation studies. Specifically, we used various encoder models for FF-CNN and KeyNet with identical TGA compensation and monogenic analysis for all abalations. Highest accuracy in detecting fractures was obtained using  6 convolutional blocks for FF-CNN, 6 convolutional blocks with a regressor for KeyNet and 6 convolutional blocks for RefineNet. Each convolutional block consisted of a convolutional layer, ReLU activation and batch normalisation. 

\section{Discussion}
We described a transporter neural network with components tailored to domain specific features seen in ultrasound. We applied our model on wrist ultrasound  which is a challenging use case due to acoustic shadowing near bony structures. Instead of using end-to-end deep learning we introduced features specific to ultrasound as a bone probability map. As a preprocessing step, we also compensated for TGA using an exponentially decaying depth dependent model. As shown in Figure \ref{fig2}, TGA compensation was able to suppress most of the  bright patches that were close to the transducer surface and highlight the bony structures located deeper in the image.

\paragraph{}
Conventional DL models in ultrasound use supervised learning which requires precisely labeled ground truth annotations. Our keypoint detection framework uses unsupervised learning and relies on information from neighbouring frames. Using the bone probability map we ensure that the neural network learns features and keypoints that are relevant in detecting wrist fractures. A key advantage of our framework is the ability to indicate the approximate location of the fracture using keypoints.

\paragraph{}
Since keypoints are detected automatically we were able to train our model with a relatively small number of videos (N $<$ 50). Our technique could be used at nearly real-time with an execution time $<$1 sec per volume using an NVIDIA V100 GPU.
\paragraph{}
Although the approach has been explained in context of wrist ultrasound, it can be extended to other anatomical structures that contain bone. For instance, it can be adapted to analyse ultrasound sequences of elbow or shoulder to detect fracture or ligament tears. With minimal modification it  could also be used for other scans where relevant artifacts can be detected using local phase. For example, lung ultrasound (LUS) images contain horizontal (A-lines) and vertical artifacts(B-lines)  that can be detected using local phase.

\paragraph{}
As future work we plan to incorporate a classification head in the transporter architecture which could be trained to detect fractures reported by human readers. We would develop a hybrid loss function which combines MSE and cross entropy for frame reconstruction and binary classification. Features learned from the generative RefineNet model could be used to initialize the new classification head, there is also the possibility of obtaining even more relevant keypoints for fracture detection by providing the FF-CNN features at the keypoint locations to the classification head with gradient backpropagation through the keypoint locations.

\paragraph{}
Our study has limitations, firstly our dataset is limited (N=30) and was collected from a single center which limits the generalizability of our results. As future work we plan a large scale multicenter study to validate the AI technique on ultrasound sweeps (which are more common in practice than 3DUS). Another limitation of our tracking technique is that we have not specifically addressed motion artifacts that are commonly seen in ultrasound scans. Although in most cases these artefacts occupy higher frequencies they could potentially affect the reconstruction of the feature map and result in incorrect keypoints. This can be addressed by associating spatial and temporal saliency features to each keypoint. Lastly, not all fractures are visible in all four views which could result in variability in ground truth on a per-patient basis. With a larger dataset we would be able to address this limitation  as we would be able to develop models specifically trained on each scan view.

\section{Conclusion}
This is the first use of an unsupervised transporter neural network to detect pathology in ultrasound. Our approach can be used in emergency care to assess 3D wrist ultrasound scans and identify relevant keypoints. We incorporated ultrasound specific features in an unsupervised learning framework and identified clinically relevant features. This automatic technique significantly reduces interobserver variability and would potentially result in more widespread use of ultrasound. Replacing X-ray examination with ultrasound would reduce radiation exposure and and improve the overall quality of emergency care.

\section{Acknowledgements}

%
%
%
%

\end{document}